\documentclass[10pt, a4paper, twocolumn, aps, pra, showpacs, longbibliography, nofootinbib,superscriptaddress]{revtex4-1}
\pdfoutput=1
\usepackage[utf8x]{inputenc}
\usepackage{ucs}
\usepackage{amsmath}
\usepackage{amsfonts}
\usepackage{amssymb}
\usepackage{makeidx}
\usepackage{cellspace,booktabs}
\usepackage{natbib}
\usepackage{lipsum}
\usepackage{bm}
\usepackage{bbm}
\usepackage{relsize}

\usepackage{hyperref}
\hypersetup{colorlinks = true, linkcolor=magenta,citecolor=blue, urlcolor=magenta, bookmarksnumbered =  true}
\usepackage{color}
\definecolor{light-gray}{gray}{0.55}

\usepackage{microtype}

\usepackage{graphicx}

\newcommand{\ssm}{\rm\scriptscriptstyle}

\newcommand{\ket}[1]{ \lvert #1 \rangle}

\begin{document}

\begin{abstract}	
The efficient validation of quantum devices is critical for emerging technological applications. In a wide class of use-cases the precise engineering of a Hamiltonian is required both for the implementation of gate-based quantum information processing as well as for reliable quantum memories. Inferring the experimentally realized Hamiltonian through a scalable number of measurements constitutes the challenging task of Hamiltonian learning. In particular, assessing the quality of the implementation of topological codes is essential for quantum error correction. Here, we introduce a neural net based approach to this challenge. We capitalize on a family of exactly solvable models to train our algorithm and generalize to a broad class of experimentally relevant sources of errors. We discuss how our algorithm scales with system size and analyze its resilience towards various noise sources.
\end{abstract}

\date{\today}
\author{Agnes Valenti}
\affiliation{Institute for Theoretical Physics, ETH Zurich, CH-8093, Switzerland}
\author{Evert van Nieuwenburg}
\affiliation{Institute for Quantum Information and Matter, Caltech, Pasadena, 91125 California, USA}
\author{Sebastian Huber}
\affiliation{Institute for Theoretical Physics, ETH Zurich, CH-8093, Switzerland}
\author{Eliska Greplova}
\affiliation{Institute for Theoretical Physics, ETH Zurich, CH-8093, Switzerland}

\title{Hamiltonian Learning for Quantum Error Correction}

\maketitle

\section{Introduction}

While finding an eigenstate $\ket{\Psi}$ of a given Hamiltonian $H$ might be a daunting task, the problem is nevertheless well defined and can often be solved by a set of existing tools. The inverse question, the identification of a Hamiltonian $H$ given the knowledge of a state $\ket{\Psi}$, is a much more complicated one and without a unique answer in the general case.

Hamiltonian engineering is one of the central objectives in the context of constructing quantum information processing and storage devices \cite{LaddJelezko2010, Preskill2018NISQ, AcinBloch2018}. Despite significant progress of quantum technologies over the last decades, the validation of the engineered Hamiltonian will always be an essential step for any quantum device. Moreover, the implemented Hamiltonian cannot be accessed directly, but only through the measurements performed on the system. Given these measurements one would like to verify that the desired Hamiltonian has been implemented correctly. In addition to that, this verification should ideally be done in an efficient and scalable way.

Reconstructing the Hamiltonian governing the systems' dynamics directly from the measurement results constitutes the inverse problem we address. Beyond not being unique, in the  quantum setting there arises an additional challenge due to the exponential size of the Hilbert space: Reconstructing even just the state alone via full quantum tomography is in general exponentially costly \cite{NielsenChuang2000} (though there are notable cases where tomography can be done efficiently \cite{TorlaiMazzola2018,CramerPlenio2010,GrossLiu2010}). However, even with one eigenstate fully reconstructed, we still do not have enough information to determine which Hamiltonian it belongs to in the general setting. 

There are two aspects one can exploit to simplify the inverse problem. First, we are typically not interested in the full Hamiltonian. For many purposes we only need to make sure we implement a Hamiltonian which possesses a ground state with desired properties, such as topological ground state degeneracies \cite{Kitaev2003}. In other words, we only aim at finding a possible parent Hamiltonian to this ground state family. Second, the implemented Hamiltonian we try to reconstruct will typically be in the vicinity of the sought after parent Hamiltonian. This property can render the reconstruction feasible and will typically not require a full tomography.

In this work, we develop a method to recover a parent Hamiltonian for a generic example important in the context of quantum devices: the family of the so-called stabilizer Hamiltonians  \cite{Gottesman1997} that lie at the heart of the theory of fault-tolerant quantum computation \cite{LidarBrun2013,Gottesman1998}. A canonical example of this family of Hamiltonians is the toric code model \cite{Kitaev2003}. The toric code is a quantum spin-$1/2$ model defined on square lattice with periodic boundary conditions. Its ground state manifold is exactly known and its structure can be used for a stable encoding of quantum states.

In particular, we design a machine learning driven method for the validation of the implementation of the toric code. Our approach addresses all of the aforementioned challenges in the solution of the inverse problem: (i) We find a minimal set of local measurements performed on a ground state that allows us to learn the system’s Hamiltonian, i.e., we don't need full tomography. (ii) Capitalizing on the proximity to the targeted toric code Hamiltonian we can train the neural network on a restricted class of {\em exactly solvable} models. Using the power of generalization of machine learning algorithms, we can then apply these trained networks to the general problem. (iii) The need for only a very restricted set of measurements and the use of efficient computing algorithms allow us to scale the solution problem beyond system sizes required for future experiments. Moreover, the generalization power of neural nets equips our approach with a certain resilience to experimental noise.

Our work has to be seen in the context of recent developments in the field. Specifically, using a trusted quantum simulator \cite{GranadeFerrie2012,WiebeGranade2014,WiebeGranade2015,WangPaesani2017}, determining a Hamiltonian from eigenstate dynamics \cite{BurgarthMaruyama2009,DiPaternostro2009,ZhangSarovar2014,DeOswald2016,VinetZhedanov2011,SoneCappellaro2017,WangDong2018}, compressed sensing \cite{RudingerJoynt2015}, RBM quantum tomography \cite{KieferovaWiebe2017}, maximum-likelihood generalization\cite{Kappen2018}, and mapping on straightforward parameter estimation problem \cite{GreplovaAndersen2017} have been used to address the inverse problem. Recently, methods to recover a local Hamiltonian based on measurements on a single eigenstate have been proposed \cite{QiRanard2017,ChertkovClark2018,GreiterSchnells2018,BaireyArad2019}. The introduced frameworks, however, come either with the necessity to measure long-range correlators or are not applicable to Hamiltonians consisting of commuting terms \cite{BaireyArad2019}. 

The ability to efficiently validate Hamiltonians with topologically protected ground state degeneracies has direct implications for quantum error correction \cite{LidarBrun2013,FoselTighineanu2018}. One way of using the toric code as a quantum memory is to prepare the ground state in order to encode logical qubits in its ground state manifold \cite{Kitaev2003}. In practice, this means designing the many-body interactions that the stabilizer Hamiltonians contains. While many-body interactions can be difficult to implement in general in an experiment, there are theoretical proposals and experimental progress towards achieving the four-body interaction needed for the toric code in a variety of platforms \cite{BravyiSuchara2014, ChancellorZohren2017, MezzacapoLamata2014, MullerHammerer2011, MizelLidar2004, GurianCheinet2012, GladchenkoOlaya2009, BarreiroMuller2011}. However, only through the reliable validation of these schemes can one lift these proposals to a potential avenue for topologically protected quantum memories. Here we provide such a validation protocol. 

In an alternative approach to using the toric code, one arrives at a ground state without the need to engineer any Hamiltonian, but an arbitrary state is successively projected into the desired state by a series of projective measurements \cite{FowlerMariantoni2012, ReiterSorensen2017}.  The performance of these methods depends on the frequency, correlations and types of errors \cite{FowlerWhiteside2012, BravyiDuclos2012, DennisKitaev2002,SwekeKesselring2018, Andreasson2018} and we comment below how our algorithm relates these approaches. 

This paper is organized as follows: In Sec. \ref{sec:two} we discuss the family of exactly solvable stabilizer Hamiltonians and the phase transitions they manifest. Sec. \ref{sec:three} addresses the neural network based Hamiltonian learning scheme we designed for this class of topological models. In Sec. \ref{sec:four} we present the numerical analysis of the convergence and errors of the model and elaborate on its resilience to experimental noise. We discuss our results presented here in the broader context of the field and their implication for the future research directions in Sec \ref{sec:five}.

\begin{figure}
\centering
\includegraphics[scale=0.27]{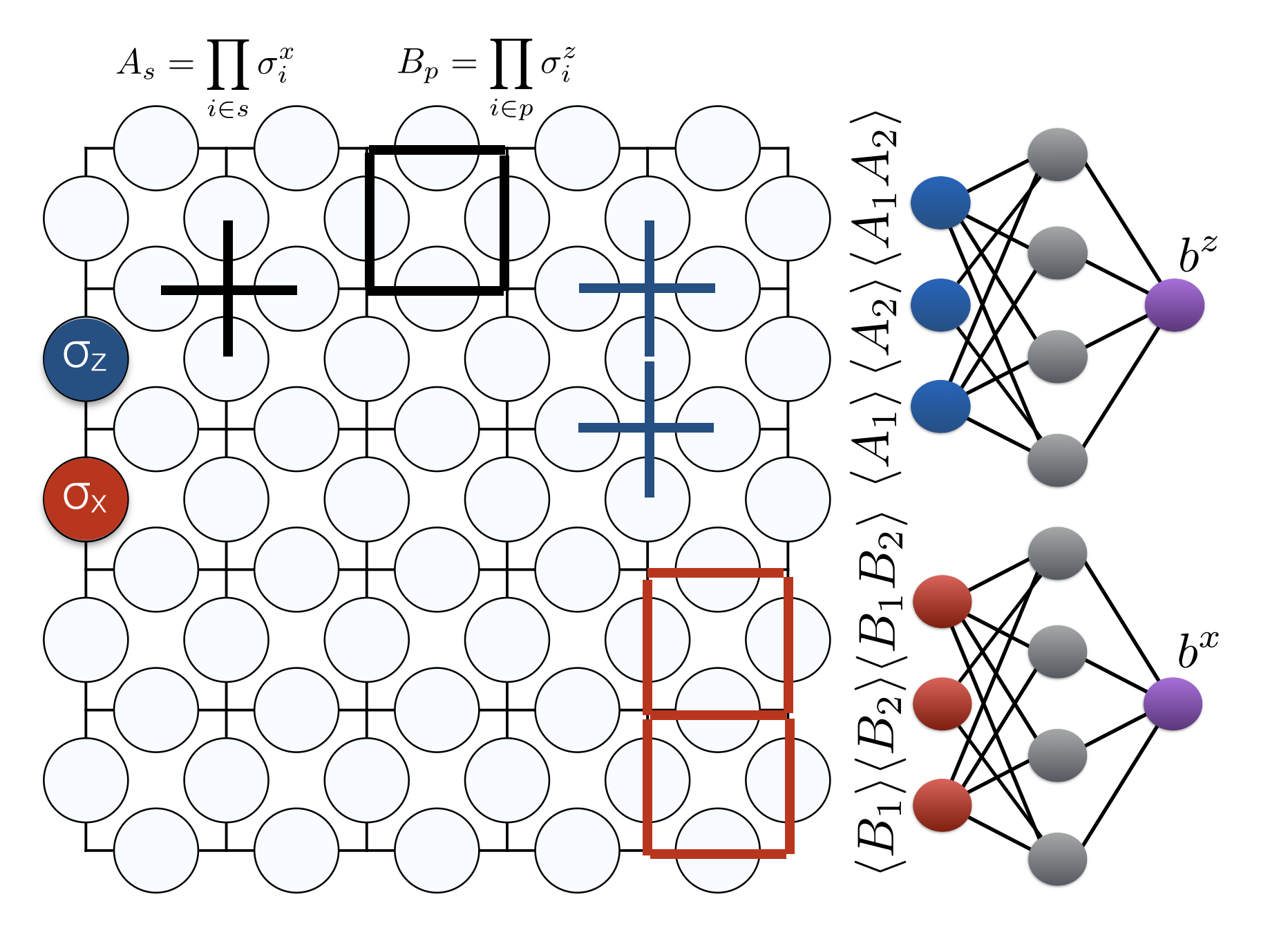}
\caption{Schematic of our method: We add background fields to the stabilizer Hamiltonian. We use expectation values of star operators, $A_i$, to determine $\sigma_z$ fields (shown in blue) and expectation values of plaquette operators, $B_i$, to determine $\sigma_x$ fields (shown in red).}
\label{fig:lattice_nn}
\end{figure}

\section{Solvable topological models}
\label{sec:two}

Kitaev's toric code model \cite{Kitaev2003} is defined on a $k\times k$ square lattice with periodic boundary conditions and spin-$1/2$ degrees of freedom located on the edges. The Hamiltonian consists of the sum of four-spin interaction terms
\begin{align}
H_{\ssm TC}= -\sum \limits_{s} A_s  -  \sum \limits_{p}B_p,
\label{eq::tcH1}
\end{align}
where the stabilizer operators are defined as 
\begin{align}
	A_s&=\prod \limits_{i \in s} \sigma_i^x\quad\mbox{and}\\
	B_p&=\prod \limits_{i \in p} \sigma_i^z.
\end{align}
Here, $s$ denotes the set of four spins around a vertex and $p$ the set of four spins around a plaquette, respectively (see Fig.~\ref{fig:lattice_nn}).
Since the stabilizer operators, $A_s$ and $B_p$, mutually commute, the ground state manifold is exactly known. It corresponds to eigenstates of all operators $A_s$ and $B_p$ with maximal eigenvalues $+1$.  

The ground state degeneracy depends on the topology of the manifold the model is defined on. On a torus, there are four degenerate ground states $|{\rm GS}_i\rangle$, $i\in \{0,1,2,3\}$. Moreover, the topological order of these ground states results in stability of the degeneracy against arbitrary local perturbations. A qubit state can then be encoded through a suitably chosen superposition
\begin{align}
|\Psi\rangle_{\rm encoded}=
\sum_{i=0}^3\alpha_i|{\rm GS}_{i}\rangle.%+\beta|{\rm GS}_{1}\rangle+\gamma|{\rm GS}_{2}\rangle+\delta|{\rm GS}_{3}\rangle.
\label{eq::Psiencoded}
\end{align}
The quantum operations on this state can be performed by so-called non-contractible loops, open strings of $\sigma^z$ or $\sigma^x$ operators accross the lattice. This property of the toric code ground state provides exceptionally stable encoding for the state of a logical qubit: A lattice-size long string of qubits needs to be flipped for the state of the logical qubit to change. For the purpose of our Hamiltonian learning we can restrict the discussion to one of the four states and we will use
\begin{equation}
|{\rm GS}\rangle_{{\ssm TC}}=\frac{1}{2}\prod \limits_{s} (1+A_s)|0\rangle,
\end{equation}
where $\ket{0}$ is a reference state of the lattice with all spins up. We comment below on why our method is insensitive to the choice made here.
\begin{figure}
\centering
\includegraphics[scale=0.8]{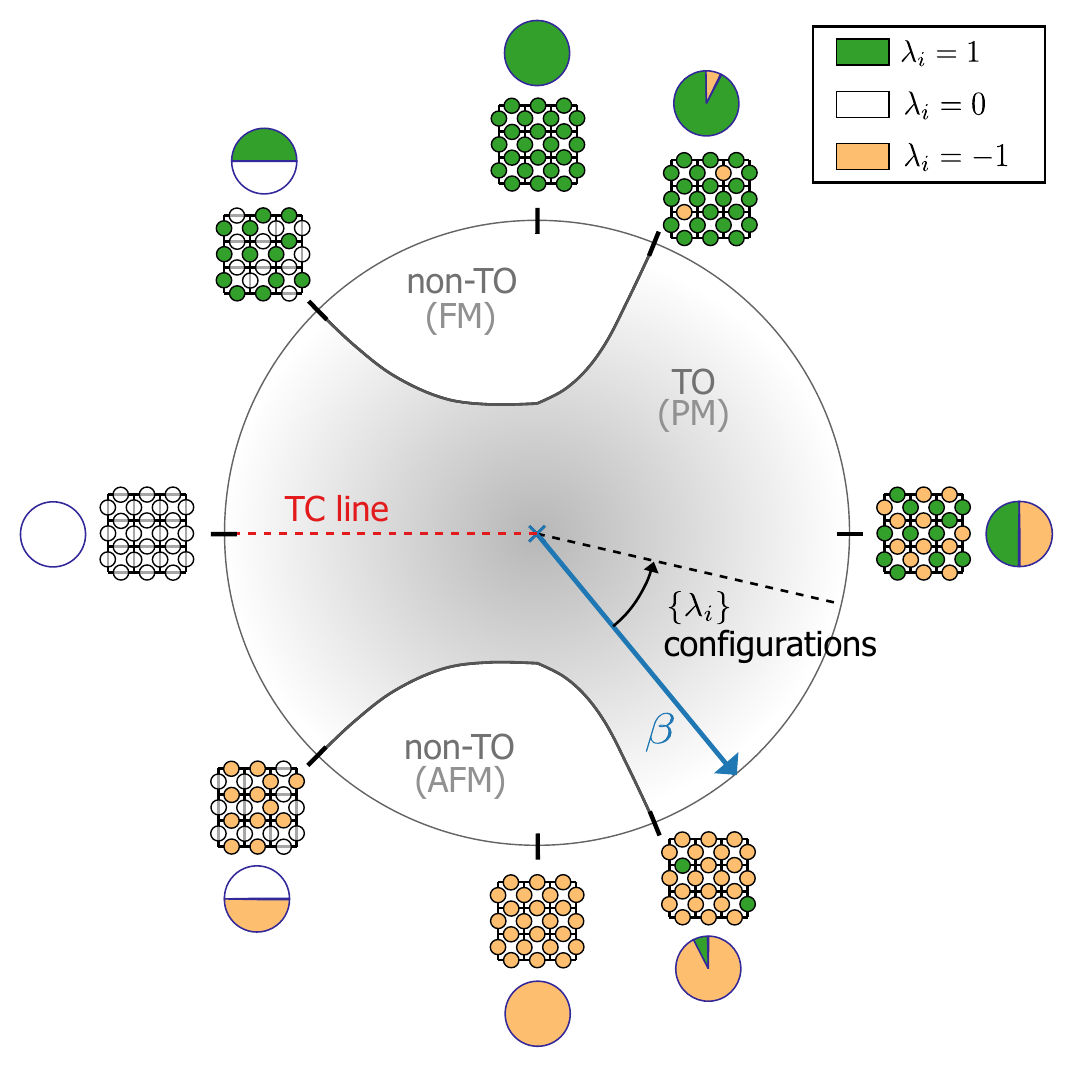}
\caption{Phase diagram of the modified toric code model. The field strength $\beta$ (blue axis) and configuration $\{\lambda_i\}$ are encoded as amplitude and angle in the phase diagram. Important field configurations are marked at the phase diagram
boundary with an example field configuration and a small diagram. The small diagrams show the percentages of fields $\lambda_i$ being equal to $+1$, $0$ or $−1$. A phase transition out of the topologically ordered phase (TO) into the non-topologically ordered phase (non-TO) occurs at the black phase boundary. The phase of the corresponding 2D classical spin model is denoted by PM (paramagnetic) , FM (ferromagnetic) or AFM (antiferromagnetic). If all fields are equal to zero, the system corresponds to the pure toric code model (\ref{eq::tcH1}). In the phase diagram, these configurations are indicated by a red dashed line with label ``TC line''.}
\label{fig:pdiag}
\end{figure}

A question can be posed about how the ground state changes if the stabilizer Hamiltonian is perturbed. In Refs. \cite{CastelnovoChamon2008, TsomokosOsborne2011} it was shown that there exists a way to leave the topologically ordered toric code phase while keeping the model analytically solvable.
In particular, we can add $\sigma^z$-terms to the star operators, $A_s$, in a way that slightly modifies the ground state, while keeping its degeneracy untouched. We write this Hamiltonian as
\begin{equation}
H=\sum \limits_{s} \left(A_s + e^{-\beta \sum \limits_{i \in s}\lambda_i\sigma_i^{z}}\right)+  \sum \limits_{p}B_p, \label{eq:hamil_beta0}
\end{equation} 
where $\lambda_i \in [-1,1]$. The positive, real-valued parameter $\beta$ drives the phase transition out of the topological phase. 

The particular position of the phase transition depends on the distribution of fields given by $\lambda_i$. The ground state of $H$ can be expressed in terms of the original toric code ground state, $\ket{{\rm GS}}_{\ssm TC}$ as 
\begin{equation}
 |\Psi\rangle = \frac{1}{\sqrt{Z}} e^{\frac{\beta}{2}\sum \limits_{i} \lambda_i \sigma_i^z}|{\rm GS}\rangle_{\ssm TC}.
 \label{eq:gs_beta}
\end{equation}
The normalization factor $Z$ is given and explained in Appendix \ref{app:normalization}. The above described modification is not the only approach for keeping the Hamiltonian \eqref{eq::tcH1} analytically solvable while driving the phase transition. For example, we can symmetrically modify the plaquette terms, $B_p$ by $\sigma^x$-terms. In particular, adding $\exp\{-\beta \sum_{i \in p}\lambda_i\sigma_i^{x}\}$ to each plaquette would yield an analogous analytical ground state, $ |\Psi\rangle =\frac{1}{\sqrt{Z}} \exp\{\frac{\beta}{2}\sum_{i} \lambda_i \sigma_i^x\}|{\rm GS}\rangle_{\ssm TC}$. 

The added perturbations are of experimental relevance, as they simplify to small $\sigma^z$ (or $\sigma^x$) background fields when Eq. (\ref{eq:hamil_beta0}) is expanded to lowest order in $\beta$. Similarly, one obtains $\sigma^z$ (or $\sigma^x$) background fields if the field configuration $\{\lambda_i\}$ is sufficiently sparse (see Appendix \ref{app:normalization}). This effect could be a result of an imperfection or systematic errors in an engineered interaction.

For the sake of Hamiltonian learning, understanding the phase transition of model (\ref{eq:hamil_beta0}) is of central importance. Given the analytical form (\ref{eq:gs_beta}), it seems obvious that from suitably chosen measurements one should be able to infer the parameters of the Hamiltonian. However, this one-to-one correspondence might be obstructed by the set of degenerate ground states. It turns out, that as long as we are in a phase adiabatically connected to the pure toric code the Hamiltonian can be recovered independently of the specific superposition of degenerate ground states.

We show the position of phase transition in the perturbed toric Hamiltonian \eqref{eq:hamil_beta0} pictorially in Fig.~\ref{fig:pdiag}. The phase transition is found via mapping to classical spin models \cite{CastelnovoChamon2008,  TsomokosOsborne2011}. In particular, if the corresponding classical model is in its paramagnetic phase we are in the  phase of the toric code. Transitions out of this quantum phase are indicated by the ordering of the classical spins. Details about this mapping are provided in App.~\ref{app:topo}. Stability against local perturbations is a known property of topological order. The phase diagram in Fig.~\ref{fig:pdiag} shows that topological order in Kitaev's toric code model is also stable with respect to a large class of non-local perturbations.

These considerations lead to an immediate application to quantum error correction. Let us now assume four-body interactions of the toric Hamiltonian \eqref{eq::tcH1} are implemented in a chosen physical system. Consequently, the system will eventually arrive to its ground state. The sources of possible errors and inaccuracies are then systematic errors and noise inherently present in the experiment. We argue that error correction can then be performed at the level of Hamiltonian learning by finding a minimal set of measurements that has to be performed on the state at hand in order to deduce the Hamiltonian. Once the Hamiltonian is exactly found it can be corrected in order to arrive at the ideal Hamiltonian $H_{\ssm TC}$.

\section{Hamiltonian Learning}
\label{sec:three}

In order to solve the inverse problem we need to identify a minimal set of measurements that allows us to infer the faulty Hamiltonian. This requirement is intrinsically bound to the question of how we deduce the Hamiltonian from such measurements. In this section we outline how we approach this problem and how we can extend our scheme to the case of Hamiltonians where we do not have access to an exact ground state.

The measurements we need to perform to reconstruct the Hamiltonian $H$ should identify the product of field distributions given by $\{\lambda_i\}$ with the parameter $\beta$. To simplify the notation, we introduce the parameters $\{b_i\}$ to denote the product, $b_i=\beta\lambda_i$. 

Using translational invariance of the underlying lattice we find that the expectation values $\langle A_i \rangle$, $\langle A_{i+1} \rangle$,$\langle A_iA_{i+1} \rangle$, where $i$ corresponds to the chosen lattice site coordinate, contain a sufficient amount of information to recover the parameter $b_i$. 

This observation implies that for each lattice site we need to evaluate correlation functions of the star operators, $A_i=\prod_{i\in s} \sigma_i^x$, ``touching'' at a given lattice site. The function that maps these expectation values on the Hamiltonian parameters is implemented via a small neural network, see Fig.~\ref{fig:lattice_nn}. For each spin we input the expectation values $\langle A_i \rangle$, $\langle A_{i+1} \rangle$,$\langle A_iA_{i+1} \rangle$ and obtain $b_i$ that determines the field strength on a given spin. 

We denote with $b^z_i$ the field strength of $\sigma_z$ fields and with $b^x_i$ the fields strength of $\sigma_x$ fields (which are determined from the expectation values of the plaquette terms $\langle B_i \rangle$, $\langle B_{i+1} \rangle$,$\langle B_iB_{i+1} \rangle$). As a consequence of symmetry under the exchange of vertices and plaquettes as well as $\sigma^x$ and $\sigma^z$, the neural network trained to identify the field parameter $b_i^z$ also succeeds in identifying the parameter $b_i^x$. 

Assuming access to many copies of the system, one can evaluate the expectation values described above. Note that the total amount of expectation values needed scales linearly with the number of parameters to be estimated and hence, linearly with the number of spins in the lattice. 

We train a small neural net on the exact ground states of the family of Hamiltonians \eqref{eq:hamil_beta0}. Owing to the analytical solution we are able to simulate lattices of almost arbitrary size. We choose a particular spin in the lattice and evaluate the expectation values $\langle A_i \rangle$, $\langle A_{i+1} \rangle$,$\langle A_iA_{i+1} \rangle$ for a range of different field distributions on the surrounding spin and a range of values of $b_i^z$.

We restrict ourselves to the field configurations $\{b_i^z\}$ such that the system is in the toric code phase, cf. Fig.~\ref{fig:pdiag}. 
Consequently, it is possible to recover the Hamiltonian $H$ even if the system is in an arbitrary state in the ground state manifold, while the training set is restricted to a specific ground state. The reason for this simplification lies in the properties of the nature of the toric code ground state: The degenerate ground states can not be distinguished by local measurements and hence, the measurement outcome of the stabilizer expectation values is identical for all states in the ground state manifold.

We use the set of expectation values $\langle A_i \rangle$, $\langle A_{i+1} \rangle$,$\langle A_iA_{i+1} \rangle$ as the input for our neural net. The labels we train the network to associate with these inputs are the field strength on the given spin $b_i^z$ (Fig.~\ref{fig:lattice_nn}). We find that a dense neural network of $3$ layers with $128$, $150$ and $128$ neurons respectively is able to reliably approximate this mapping. We detail the architecture and details of the training in App.~\ref{app:network}.

We evaluate the performance of the network on states that lie outside of manifold of ground states of exactly solvable Hamiltonians. This strategy would in principle allow us to apply the trained models on measurements of a large quantum computer or simulator. Above, we introduced two examples of toric-code Hamiltonian modifications, one through $\sigma^z$ and one through $\sigma^x$ fields
\begin{align}
H=&\sum \limits_{s} \left(-A_s + e^{- \sum \limits_{i \in s}b^z_i\sigma_i^{z}}\right)+ \nonumber \\  &\sum \limits_{p}\left(-B_p+e^{-\sum \limits_{i \in p}b^x_i\sigma_i^{x}}\right). 
\label{eq:hamil_beta}
\end{align}
Separately, these modifications do not violate exact solvability of the model. However, in the small-$\beta$ limit this Hamiltonian covers a very general set of errors that may occur in an actual quantum device. Therefore, we showcase our method on this Hamiltonian.

The above Hamiltonian cannot be solved for large systems. However, we can use the network trained on the solvable Hamiltonian. We then apply it to the ground state of (\ref{eq:hamil_beta}) which for small system sizes can be obtained by brute force diagonalization.
\begin{figure}
\centering
\includegraphics[scale=1]{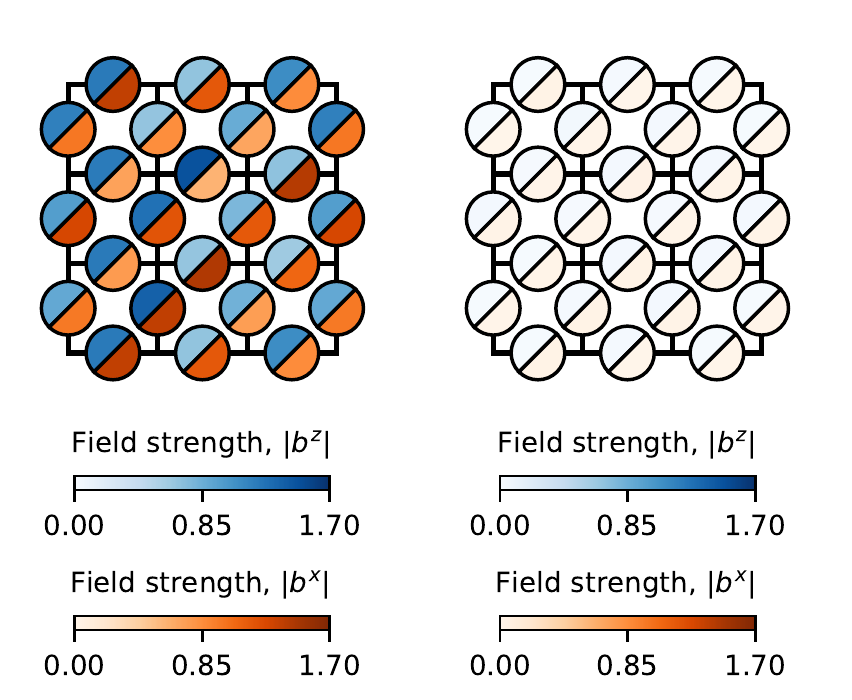}
\caption{Visualization of the removal of the fields for a small lattice with arbitrary fields. The left panel shows a possible field distribution in Hamiltonian \eqref{eq:hamil_beta}. The right panel shows the field distribution after five iterations of our protocol are performed. This illustrates the arrival to the toric code Hamiltonian \eqref{eq::tcH1}.}
\label{fig:field_small}
\end{figure}

Let us introduce an iterative algorithm designed to evaluate the fields that characterize Hamiltonian (\ref{eq:hamil_beta}). We take one of the above described ground states and evaluate the required correlators $\langle A_i \rangle$, $\langle A_{i+1} \rangle$,$\langle A_iA_{i+1} \rangle$ and $\langle B_i \rangle$, $\langle B_{i+1} \rangle$,$\langle B_iB_{i+1} \rangle$ for each of the spins, and we use them as input to the trained neural network. For each spin we obtain the magnitude of the field strength in $\sigma^x$ and $\sigma^z$ direction. We follow this step by numerically removing the found fields from the original input Hamiltonian. While in a numerical simulation this is a straightforward task, in a quantum simulation this would be implemented by adjusting the interaction parameters to cancel the fields and re-initializing, or by an adiabatic transition between two Hamiltonians. An implementation of the corresponding set of gates is also possible. We use the ``corrected'' state as the new input for the same neural network, remove the resulting fields, and repeat the procedure iteratively until convergence. The iterative application is not completely straightforward as $\beta$ and $-\beta$ result in the same expectation values of $A_s$ and $B_p$. Hence, the iterative procedure includes a decision tree for the sign of the field. With these adjustments (addressed in detail in Appendix \ref{app:network}) the correct fields (and therefore the correct Hamiltonian) are found within five iterations. We illustrated the algorithmic field removal in Fig.~\ref{fig:field_small}.

\section{Results}
\label{sec:four}

We define two quantities to characterize the performance of our method. First, we introduce a measure at the level of the resulting corrected quantum ground states: The probability that physical qubits pertain spin flip or phase errors after the projection onto the stabilizers.  Second, we put forward a criterion at the level of the corrected Hamiltonian. This measure is simply a distance between the ideal and the corrected Hamiltonian in a suitably chosen metric.

To formulate the error measure for the corrected states we draw the connection to standard error correction strategies: One starts from an {\em arbitrary quantum state} and implements a projection on the eigenspaces of the operators $A_s=\prod_{i\in s} \sigma_i^x$, $B_p=\prod_{i\in p} \sigma_i^z$ by measuring the state of ancillary qubits that are entangled with star and plaquette operators \cite{FowlerMariantoni2012}. This projection results in an eigenstate of Hamiltonian \eqref{eq::tcH1}. Using the measurement outcomes, a decoder identifies single spin operations that map the state to the ground state of Hamiltonian \eqref{eq::tcH1}. The decoder will succeed if the probability of spin and phase flip errors between the projective measurements is lower than a given threshold.
This threshold ranges anywhere between $2\%$-$11\%$ depending on the chosen error model \cite{FowlerWhiteside2012, BravyiDuclos2012, DennisKitaev2002, SwekeKesselring2018, BreuckmannNi2018, Ni2018, TorlaiMelko2017, Andreasson2018}.

For our case we do not start from an arbitrary state, but from the {\em ground state of the corrected parent Hamiltonian}. We can relate the precision of our method to standard decoding techniques by calculating the probability that a single physical qubit will flip its spin or phase when projecting into the stabilizer eigenspace. The calculation of a single qubit spin or phase flip can be deduced from the stabilizer expectation values $\langle A_s \rangle$ and $\langle B_p \rangle$ and is detailed in App. \ref{app:errormeasures}.
 
\begin{figure}
\centering
\includegraphics{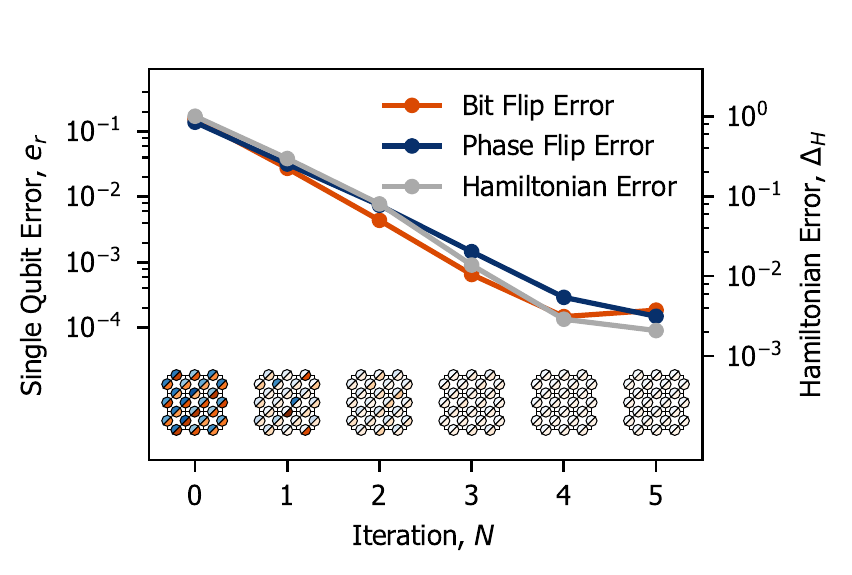}
\caption{Error probability for bit and phase flips (red and blue respectively, left hand axis) and Hamiltonian error (grey, right hand axis) as a function of the iteration of the protocol, see text for details. At each iteration step the field configuration strength is shown (color scheme same as in Fig.~\ref{fig:field_small}).}
\label{fig:state_corr}
\end{figure}

In Fig.~\ref{fig:state_corr}, we show the probability of a single bit or phase flip error as a function of the iteration of our algorithm. The iteration axis also contains pictorial illustrations of the absolute field strengths on the state. When projecting on the stabilizer eigenstates from the original faulty eigenstate we find errors well above the decoding thresholds (typically above $12\%$). We are, thus, able to show that even a small systematic error present in the Hamiltonian engineering can potentially confuse a decoder.  After five iterations we arrive to a single qubit error probability of order of $10^{-2}\%$.\footnote{This translates into the probability for wrong stabilizer measurements of the same order $~10^{-2}\%$.} In other words, if we attempted to decode the initial faulty ground state, the decoder may fail. After our correction procedure it is always highly likely to succeed. 
 
Another measure that quantifies the precision of Hamiltonian learning can be implemented on the level of the Hamiltonian directly \cite{BaireyArad2019}. This method relies on expanding the family of Hamiltonians we would like to estimate in a suitable basis and then measure the distance of the estimated coefficients from the ideal ones. In particular, the Hamiltonian is expressed as $H=\sum_{m} c_m S_m$, where $\{S_m\}$ is an operator basis with expansion coefficients $c_m$. Once we make sure that the chosen basis $\{S_m\}$ is independent of the parameters to be estimated, the Hamiltonian error simply translates into the distance
\begin{align}
\label{eq::delta2}
\Delta_H=||\hat{c}_{\ssm true}-\hat{c}_{\ssm recovered}||_2,
\end{align}
where $\hat{c}_{\ssm true}$ and $\hat{c}_{\ssm recovered}$ are the normalized vectors of exact and found coefficients respectively. In our case $\hat{c}_{\ssm true/recovered}$ are non-linear functions of $b_i=\beta\lambda_i$, the parameters the neural network is estimating. We show this functional dependence together with the expansion of the Hamiltonian in App.~\ref{app:errormeasures}. The resulting Hamiltonian error is shown as a function of iterations of the protocol in Fig.~\ref{fig:state_corr}. We normalize the Hamiltonian error such that the maximal value is one. After five iterations of the protocol, the distance between expansion coefficients decreases to order of $10^{-3}$.

\begin{figure}
\centering
\includegraphics{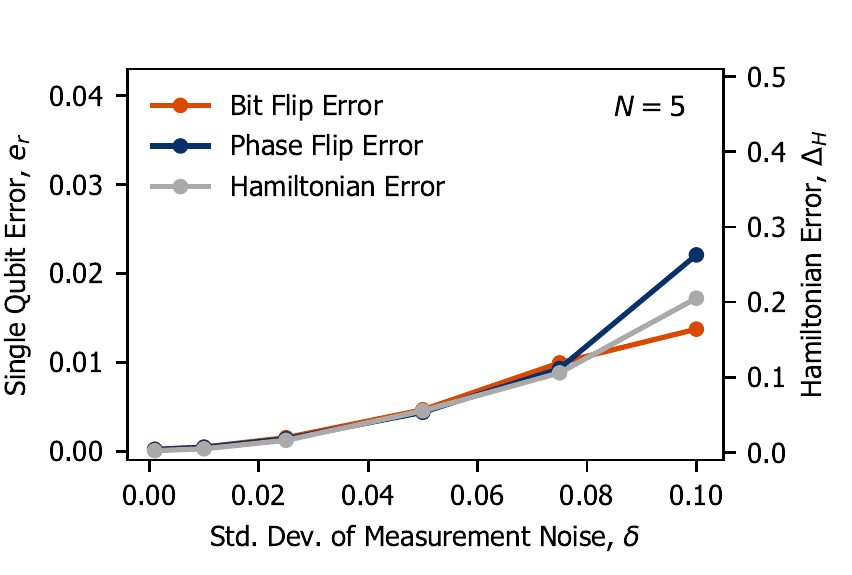}
\caption{Single qubit error probability and Hamiltonian error as a function of a standard deviation of the measurement noise. In red we show the bit flip error, in blue the phase error and in grey the Hamiltonian error. The Hamiltonian error is plotted on the right hand axis. The results are shown after $N=5$ iterations of our protocol.}
\label{fig:error+hamil+noise}
\end{figure}

Both state and Hamiltonian error explained up to this point were evaluated under the assumption of perfect experimental readout. We now briefly discuss how our method performs when this assumption is relaxed. We assume that the experimental input for the neural network is in the form of the measured expectation values. Experimental noise would then enter through these expectation values not being evaluated correctly. Additionally, estimating an expectation value using a finite number of samples induces a statistical uncertainty.
We simulated this scenario by adding Gaussian noise on our numerically evaluated inputs $\langle A_s \rangle$, $\langle B_p \rangle$. We show the behavior of the single qubit errors as well as the Hamiltonian error in Fig.~\ref{fig:error+hamil+noise}. We observe that even in presence of Gaussian noise the single qubit error rate after projection onto the stabilizers is reliably reduced below $5\%$.

\begin{figure}
\centering
\includegraphics{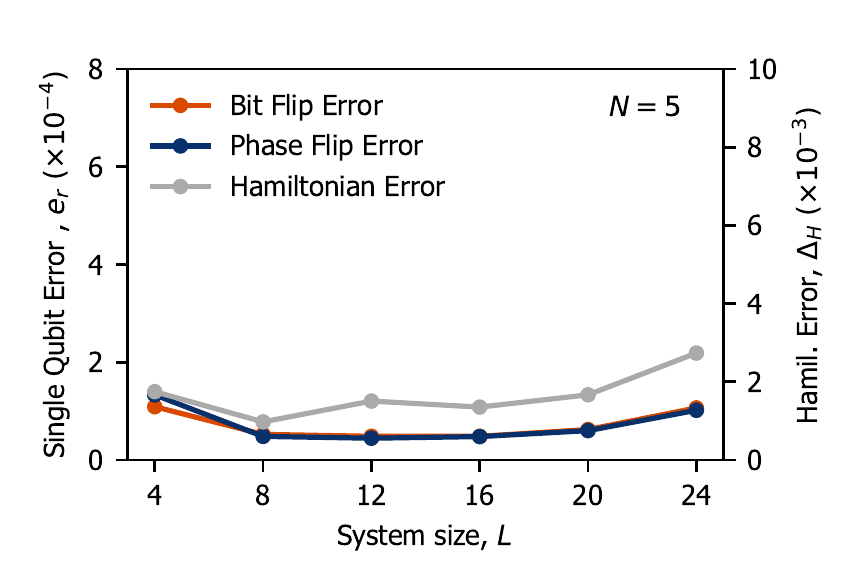}
\caption{Single qubit error probability and Hamiltonian error as a function of the lattice size. The used notation is the same as in Fig.~\ref{fig:error+hamil+noise}.}
\label{fig:error+hamil}
\end{figure}

All simulations so far are conducted for a lattice with $18$ spins, as a full quantum simulation is required to simulate the correction procedure for the non-solvable model [see Eq. (\ref{eq:hamil_beta})]. However, this restriction is not necessary when it comes to training of the neural network. In an experiment, the presented Hamiltonian learning procedure can therefore be applied to almost arbitrary lattice sizes. We demonstrate the scaling behavior by training the network separately for different lattice sizes while the size of the training set is fixed. The computation time to generate the training set via Monte Carlo sampling only increases linearly with number of qubits, as the number of sweeps through the lattice is kept constant. The correction process can only be simulated by keeping the model solvable. We want to emphasize that this restriction appears purely for testing purposes and is not of relevance in an experimental setup.
In Fig.~\ref{fig:error+hamil} the results of our simulations are shown, i.e. the error measure outcomes are independent of lattice size.

\section{Discussion}
\label{sec:five}

In this work we have introduced a machine-learning driven method for scalable quantum Hamiltonian learning based on a toy model relevant for quantum information processing, the toric code. Scalability was achieved by training the neural network on exactly solvable models. We have shown that our method performs well even for states that lie outside this class of analytical solutions. Using knowledge of topological phases and the stability of the toric code to a large class of non-local perturbations, we justified the restriction of the correction to the topological phase. However, an expansion of the correction process outside of the topological phase should be a straightforward procedure. 

Our work is complementary to standard approaches to quantum error correction. More specifically, the performance of a standard decoder is significantly improved when combined with the procedure introduced here. When the Hamiltonian is engineered precisely, logical errors are suppressed. We provided a tool for precise engineering based on measuring a minimal set of local expectation values scaling linearly with number of qubits. Implementing a practical protocol that incorporates both the Hamiltonian learning techniques presented here and stabilizer measurement based decoders would therefore be a natural step towards running a fault-tolerant quantum computations.

\section*{Acknowledgements}
We acknowledge discussions with Aleksander Kubica, Tomas Jochym O’Connor, Gil Refael, Netanel Lindner, Christoph Bruder and Mohammad Hafezi.
We are grateful for financial support from the Swiss National Science Foundation, the NCCR QSIT. This work has received funding from the European Research Council under grant agreement no. 771503. This research was supported in part by the National Science Foundation under Grant No. NSF PHY-1748958 and by the Swiss National Science Foundation under Grant No. 183945. AV acknowledges financial support of the ETH Master Scholarship Programme.

\begin{appendix}

\section{Supplementary Calculations}
\label{app:normalization}
The ground state structure in limiting cases of the disordered toric code model introduced throughout this work is detailed in this section.

\subsection{Normalization factor}
The ground state of the disordered Hamiltonian defined in Eq. (\ref{eq:hamil_beta0}) is of the form
\begin{align}
|\Psi\rangle = \frac{1}{\sqrt{Z}} e^{\frac{\beta}{2}\sum \limits_{i} \lambda_i \sigma_i^z}|{\rm GS}\rangle_{\ssm TC}.
\end{align}
To explain the normalization factor $Z$, we re-write the toric code ground state as
\begin{align}
\label{eq::gsTCneu1}
|{\rm GS}\rangle_{\ssm TC}&=\frac{1}{2}\prod \limits_{s} (1+A_s)|0\rangle \\
\label{eq::gsTCneu2}
&=\frac{1}{2} \mathop{\sum_{n_i=0,1}} \limits_{i=1,...k^2} \mathop{\prod_{i=1}^{k^2}} A_{s_i}^{n_i} |0\rangle \\
\label{eq::gsTCneu3}
 &=\sum \limits_{g\in G} g |0\rangle,
 \end{align}
where $G$ denotes the abelian group whose elements $g$ are all possible spin-flip operations defined by the action of products of vertex operators on an initial spin-configuration \cite{CastelnovoChamon2008}. In the toric code, periodic boundary conditions in all directions are imposed leading to the relation $\prod_s A_s={1}$. The group elements of the group $G$ are only determined {\it modulo} the factor $\prod_s A_s$. As a consequence, the number of elements in the group is equal to half the number of possible products of vertex operators.
We can re-write the ground state $|\Psi\rangle$ using the introduced notation
\begin{align}
\label{eq::PsiM}
|\Psi\rangle = \frac{1}{\sqrt{Z}}\sum \limits_{g\in G} e^{\frac{\beta}{2}\sum \limits_{i} \lambda_i \sigma_i^z(g)}g |0\rangle.
\end{align}
Here, $\sigma_i^z(g)$ can take the values $\pm 1$. More concretely, $\sigma_i^z(g)$ corresponds to the eigenvalue of the operator $\sigma_i^z$ on the eigenstate $g |0\rangle$. The normalization factor (partition function) has the following form
\begin{align}
Z:=\sum_{g \in G} e^{\beta\sum \limits_{i} \lambda_i \sigma_i^z(g)}.
\label{eq:app_partition}
\end{align}

\subsection{Disorder in the limit of small fields}
In particular cases, the disorder defined in Eq. (\ref{eq:hamil_beta0}) simplifies to $\sigma^z$ ($\sigma^x$) single-spin fields. In the limiting case of small fields ($\beta \ll 1$), we expand the added term as
\begin{align}
 e^{-\beta \sum \limits_{i \in s}\lambda_i\sigma_i^{z}} \approx 1-\beta\sum \limits_{i \in s}\lambda_i\sigma_i^{z}.
\end{align}
Terms of higher order containing $\beta^2$, $\beta^3$ and $\beta^4$ are neglected as a result of the approximation. Hence, we arrive at a model containing only $\sigma^z$ fields acting on single spins, and interactions between spins vanish. More specifically, the Hamiltonian of the disordered model in this limit is given by \cite{CastelnovoChamon2008}
\begin{align}
H \approx H_{\text{TC}}-    
2\beta\mathop{\sum_{i}} \lambda_{i} \sigma_{i}^{z}+{\rm const.}, \ \ \ \ \beta \ll 1.
\end{align}

\subsection{Disorder in the limit of sparse fields}
A similar approximation can be made for sufficiently sparse fields. We denote with ``sufficiently sparse'' here a field configuration, where at most one field parameter $\lambda_i$ per vertex is not equal to zero. If this condition is met, all interactions between spins vanish and only fields acting on single spins remain
\begin{align}
\label{eq::exponential8}
e^{-\beta \sum \limits_{i \in s}\lambda_i\sigma_i^{z}} &=\cosh(\beta \lambda_{s_0}))1-\sinh(\beta \lambda_{s_0}) \sigma_{s_0}^{z}.
\end{align}
The index $s_0$ denotes the spin in vertex $s$ for which $\lambda_{s_0} \neq 0$. 
 We insert this simplification into the Hamiltonian defined in Eq. (\ref{eq:hamil_beta0}) and arrive at
\begin{align}
&H_{\rm sparse} =H_{\text{TC}}-    
\mathop{\sum_{i \in U}}2\sinh(\beta \lambda_{i}) \sigma_{i}^{z} +{\rm const.}
\label{eq::HamiltonianMsparse}
\end{align}
Here, $U$ is defined as the sparse set of spins $i$ with field strength $\lambda_i \neq 0$.
We conclude, that we arrive at a model consisting of the toric code model with $\sigma_z$ fields and no additional interactions.

\section{Topological phases}
\label{app:topo}

\begin{figure}[btp]
\centering\includegraphics[width=5cm]{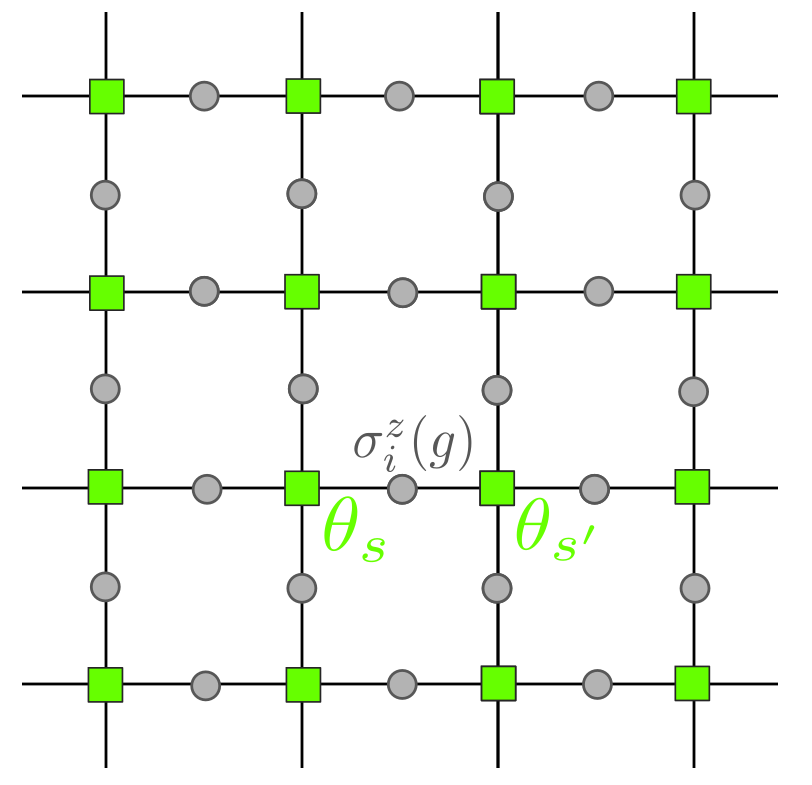}
\caption{The mapping to a classical (pseudo-)spin model. The introduced pseudo-spins on the lattice sites are marked in green, the spins (qubits) on the edges in gray. The mapping is given by the relation $\sigma_i^z(g)=\theta_{s}\theta_{s'}$ for a spin and two adjacent pseudo-spins.}
\label{fig::Isingmapping}
\end{figure}

We analyze topological order and phase transitions of the disordered toric code model in this section. For this purpose, we map the system to a classical spin model as done in \cite{CastelnovoChamon2008}. 
The ground state of the modified toric code \eqref{eq:hamil_beta0} is given by [Eq. (\ref{eq::PsiM})]
\begin{align}
|\Psi\rangle = \frac{1}{\sqrt{Z}}\sum \limits_{g\in G} e^{\frac{\beta}{2}\sum \limits_{i} \lambda_i \sigma_i^z(g)}g |0\rangle.
\end{align}
It should be noted, that the group element $g$ is only determined by the product $\prod_{s \in S_i} A_s$ {\it modulo} the product of all vertex operators, due to periodic boundary conditions. 
We can map every product of vertex operators to a pseudo-spin configuration, defined as follows: Artificial degrees of freedom  $\theta_s \in \{-1,1\}$ are introduced on all vertices $s$.
The value of $\theta_s$ determines, if the vertex $s$ is flipped (contained in the product of vertex operators we map the pseudo-spin configuration to). More specifically, the vertex $s$ is flipped if $\theta_s=-1$.
We can now make a change of variables. The eigenvalue $\sigma_i^z(g)$ of the spin $i$ depends on the two adjacent vertices $s, s'$ being flipped or not. As a consequence, this eigenvalue is given by $\sigma_i^z(g)=\theta_s \theta_s'$, resulting in a mapping between the the spin configuration $\{\sigma_i^z(g)\}$ and the pseudo-spin configuration $\{\theta_s\}$ with a gauge freedom of an overall factor of $(-1)$. The mapping is further illustrated in Fig.~\ref{fig::Isingmapping}.

Let us define an Ising model on the pseudo-spin configuration with the Hamiltonian given by \cite{CastelnovoChamon2008, TsomokosOsborne2011}
\begin{align}
H_{\rm Ising}=-\sum_{\langle s, s'\rangle} J_{s,s'}\Theta_s\Theta_{s'}.
\label{eq::ZIsing}
\end{align}
The connection to the field parameters $\beta$ and $\{\lambda_i\}$ is given by the coupling constant $J_{s,s'}$. In particular, $J_{s,s'}/(k_B T):=\beta\lambda_i$ where $k_B T$ is a product between the Boltzmann constant $k_B$ and the temperature $T$. The edge with adjacent sites $s$ and $s'$ is denoted by the index $i$.
An indication of a connection between the phase transition of the quantum model and the classical spin model can be obtained from the partition function of the defined Ising model
\begin{align}
Z_{\rm Ising}&=\sum_{\{\Theta_s\}} e^{\frac{1}{k_B T}\sum_{\langle s, s'\rangle}J_{s,s'}\Theta_s\Theta_{s'}} \\
&=\sum_{\{\Theta_s\}} e^{\beta\sum_{\langle s, s'\rangle}J_{s,s'}\Theta_s\Theta_{s'}}.
\end{align}
By comparing to the partition function of the disordered toric code \eqref{eq:app_partition} model we observe that $Z_{\rm Ising}=2Z$ \cite{AbastoHamma2008}.

We can show that the phase transition of the disordered toric code is mapped to the phase transition of the Ising model by considering a well-known measure to detect quantum phase transitions, the fidelity susceptibility \cite{YouLi2007}.
The fidelity susceptibility is defined as
\begin{align}
\chi_F=-\frac{\partial^2 \ln \langle \rm \Psi (\beta) |\rm \Psi (\beta+\Delta\beta) \rangle}{\partial (\Delta\beta)^2}\bigg|_{\Delta\beta =0}.
\label{eq::fidsusceptibility}
\end{align}
Here, the state $|\Psi (\beta)\rangle$ is a ground state of a given Hamiltonian depending on the parameter $\beta$, for our case the state is defined in Eq. (\ref{eq:gs_beta}).
A quantum phase transition is indicated via a maximum or divergence in the fidelity susceptibility $\chi_F$ \cite{VenutiZanardi2007, GuLin2009}. In particular, this property has been shown for generic second-order symmetry-breaking quantum phase transitions \cite{ZanardiPaunkovic2006}. It has further been suggested by numerical studies \citep{AbastoHamma2008}, that the fidelity susceptibility is also able to detect a topological phase transition. The fidelity susceptibility for the introduced disordered toric code model is calculated as
\begin{align}
&\chi_F=\frac{1}{4}\frac{\sum \limits_{g \in G}(\sum \limits_{i} \lambda_i \sigma_i^z(g))^2 e^{\beta\sum \limits_{i}\lambda_i\sigma_i^z(g)}\cdot Z_{z,0}}{Z_{z,0}^2} \\
&\ \ \ \ \ \ -\frac{1}{4}\frac{(\sum \limits_{g \in G}(\sum \limits_{i} \lambda_i \sigma_i^z(g)) e^{\beta\sum \limits_{i}\lambda_i\sigma_i^z(g)})^2}{Z_{z,0}^2},
\end{align}

We can understand the connection between the phase transition of the quantum toric code model and the phase transition of the classical spin model by examining the heat capacity of the mapped Ising model
\begin{align}
C_v&=\beta^2\frac{\partial^2 \ln Z_{\rm Ising}}{\partial \beta^2}  \\
&=\beta^2\frac{\partial^2}{\partial \beta^2} \ln (2 \sum \limits_{g} e^{\beta\sum \limits_{i}\lambda_i\sigma_i^z(g)})=4\beta^2\cdot\chi_F.
\end{align}
We observe, that the heat capacity is proportional to the fidelity susceptibility of the disordered toric code model.
A similar relation has already been found in \cite{AbastoHamma2008} for the fidelity metric, an equivalent measure to the fidelity susceptibility regarding the characterization of quantum phase transitions. For the disordered toric code model, both fidelity measures simplify to the same expression.
As both the heat capacity and the fidelity susceptibility indicate a phase transition via a maximum or divergence respectively, we conclude that a phase transition in the classical spin model at a critical temperature $T=T_c$ indicates a quantum phase transition in the disordered toric code model at critical field strength $\beta_c=\frac{1}{k_B T_c}$.

We now understand, that a phase transition in the defined Ising model indicates a quantum phase transition of the disordered toric code model. It remains to examine, if the indicated quantum phase transition is topological, i.e. if the transition is made to a topologically trivial phase. This question has already been discussed in \cite{CastelnovoChamon2008, TsomokosOsborne2011}. In particular, it has been shown that the paramagnetic phase of the Ising model maps to the topological phase of the perturbed toric code model, whereas the ferromagnetic phase maps to the topologically trivial phase. The antiferromagnetic phase also maps to the topologically trivial phase due to sublattice symmetry. The argument for the mapping of Ising phases to topological phases holds only deep in the corresponding phases in the case of arbitrary field strengths. Combined with the discussion of the fidelity susceptibility, we can conclude that it also holds close to the phase transition and that the topological phase transition is sharp.

Let us examine, how the field configuration $\{\lambda_i\}$ influences the position of the phase transition at critical field strength $\beta_c$. A complete description is in general a daunting task due to the amount of degrees of freedom related to the field configuration $\{\lambda_i\}$. However, we can gain insights about the stability of the topological phase when focusing on a more conceptual $\{\lambda_i\}$-$\beta$ phase diagram (see Fig.~\ref{fig:pdiag}). In particular, we aim to show for which field configurations phase transitions out of the topological phase occur at all.

\begin{figure}[t] 
    \centering
    \includegraphics[scale=1.2]{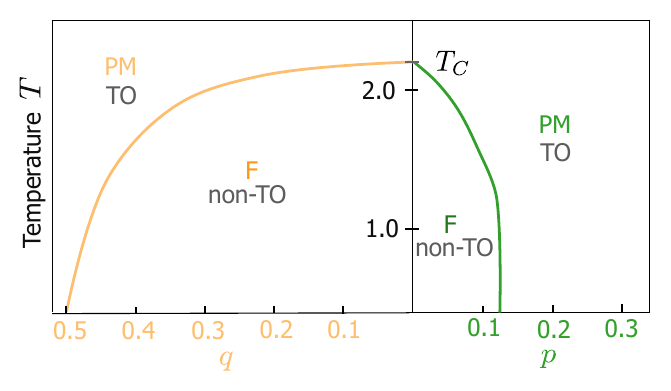}
    \caption{Phase diagram of the two probabilistic classical spin models. Temperature $T$ on the $y$-axis is plotted against dilution probabilities $q$ and $p$. The ferromagnetic (F) and paramagnetic (PM) phase and the correspondent topological order (TO) are marked for each model. The Boltzmann constant $k_B$ is set to $1$.}
\label{fig::phasediagramIsing}
\end{figure}

In order to analyze the relation between field configuration and phase transition, we examine two limiting cases. We start with a configuration with $\lambda_i=1$ $\forall i$. This configuration is mapped to an Ising model with equal bond strengths, for which the phase transition occurs at critical field strength $\beta_c \approx 0.44$. Let us ask a question whether we can modify the field configuration such that no phase transition occurs anymore.
As long as all fields are strictly greater than $0$, there can always be found a critical field strength such that the system leaves the topological phase. More concretely, the value of $\beta$ can always be chosen large enough such that every product of fields $\beta\lambda_i$ is larger than $\beta_c\approx 0.44$.  Therefore, a phase transition must occur.
Due to sublattice symmetry, the same statement holds for all fields strictly smaller than $0$.

There are two options to arrive at a configuration where no phase transition occurs. This can be achieved by obtaining a configuration of neither only strictly positive nor only strictly negative fields. One option lies in setting a fraction of fields equal to $0$ and arriving at a mixture of strictly positive fields and fields equal to $0$ (or strictly negative fields and fields equal to $0$). Here, we can examine the critical concentration of fields, that have to be removed such that no phase transition occurs. In other words, we are interested how sparse the configuration has to be such that the topological phase is never left. To answer this question, we consider a well-known probabilistic classical spin model: an Ising model with random bond dilution.
In particular, the Hamiltonian of the diluted Ising model is given by
\begin{align}
&H=-\sum \limits_{\langle s,s'\rangle} J_{s,s'} \theta_s \theta_s', \\ 
&P(J_{s,s'})=q\delta(J_{s,s'})+(1-q)\delta(J_{s,s'}-1),
\label{eq::Isingbonddilution0}
\end{align}
corresponding to
\begin{align}
\beta=\frac{1}{k_B T}, \ \
P(\lambda_i)=q\delta(J_{s,s'})+(1-q)\delta(J_{s,s'}-1). \nonumber
\end{align}
Here, $P(J_{s,s'})$ is the probability distribution of a bond strength, translating into the probability distribution $P(\lambda_i)$ of a field $\lambda_i$ with $i$ denoting the edge connecting the sites $s$ and $s'$.
The question of the critical fraction of fields set to zero hence corresponds to a critical dilution probability $q$. The model has been well-studied \cite{HoneckerPicco2001, PiccoHonecker2006, RegerZippelius1986, GruzbergRead2001} and the critical concentration found to be $q_c=0.5$. In other words, the presence of the background field on less than half of the spin cannot drive the system out of the topological phase regardless of the field strength.

Another way one can modify the field configuration in order to preserve topological order is to flip signs of a fraction of fields. We can make use of a probabilistic Ising model, which has already been studied in connection to the disordered toric code model \cite{TsomokosOsborne2011}. In this model, we consider a bond dilution determined by the probability distribution
\begin{align}
P(J_{s,s'})=p\delta(J_{s,s'}+1)+(1-p)\delta(J_{s,s'}-1),
\label{eq::Isingbonddilution1}
\end{align}
where $p$ is the probability that the sign of a bond strength is flipped. The critical value of $p$ has been found to be  $p_c \approx 0.12$ \cite{HoneckerPicco2001, PiccoHonecker2006, RegerZippelius1986, GruzbergRead2001}. Due to sublattice symmetry, the phase diagram for this model is symmetric with respect to $p=0.5$. Hence, a second critical value occurs at $p_c=0.88$. The $J-T$ phase diagram of both models is well-known and shown in Fig.~\ref{fig::phasediagramIsing}.

The found relations between field configuration and phase transition are summarized into a conceptual phase diagram, shown in Fig.~\ref{fig:pdiag}.
The field configuration $\{\lambda_i\}$ and the field strength $\beta$ are encoded as angle and amplitude in the phase diagram. The field distribution $\{\lambda_i\}$ is varied continuously with change of angle in between the configurations explicitly marked in the diagram. The depicted configurations either mark a phase transition or correspond to a limiting case of field strengths. An exemplary field distribution and a small diagram indicating the percentages of occurring field strengths illustrate the depicted configurations.
The field configurations on the left hand side do not contain fields of different signs simultaneously. On the right hand side, all fields are different from $0$.
The phase diagram is symmetric with respect to its horizontal axis as a result of sublattice symmetry.

\section{Neural Network}
\label{app:network}

\subsection{Network training and architecture}
The architecture and training process of the artificial neural network used for the Hamiltonian learning process is detailed in this section.
We have introduced in the main text a class of Hamiltonians of the form
\begin{align}
H=&\sum \limits_{s} \left(-A_s + e^{- \sum \limits_{i \in s}b^z_i\sigma_i^{z}}\right)+ \nonumber \\  &\sum \limits_{p}\left(-B_p+e^{-\sum \limits_{i \in p}b^x_i\sigma_i^{x}}\right).
\end{align}
Our algorithm aims at determination of the parameters $\{b^z_i\}$ and $\{b^x_i\}$.

We found the measurement set used as an input for the network heuristically. This set is obtained by minimizing the amount of measurements containing sufficient information to determine the Hamiltonian parameters. The result of this search is the set of expectation values $\langle A_s \rangle$, $\langle A_{s'} \rangle$ and $\langle A_s  A_{s'} \rangle$, that shows to be sufficient to determine the fields $|b^z_i|$.
Symmetry leads to the equivalent set of plaquette operators, $\langle B_s \rangle$, $\langle B_{s'} \rangle$ and $\langle B_s  B_{s'} \rangle$ determining fields $|b^x_i|$. 
As a result, when the trained network receives as input the vertex stabilizer expectation values, it outputs the field $|b^z_i|$ on the qubit $i$ with adjacent vertices $s,s'$. Analogously, when given the inputs $\langle B_p \rangle$, $\langle B_{p'} \rangle$ and $\langle B_p  B_{p'} \rangle$, it returns $|b^x_i|$ for qubit $i$ with adjacent plaquetter $p$ and $p'$.
We can combine the ability of neural networks to generalize knowledge with the symmetry of the model with respect to exchange of vertices and plaquettes. It follows that it is sufficient to train on the {\it solvable} model containing only fields in one direction, and use the network for evaluation of the general non-solvable model. Even though the training is restricted to either vertex stabilizer expectation values or plaquette expectation values, the symmetry ensures that both types of expectation values can be taken as input for evaluation and consequently both field configurations $\{b^z\}$ and $\{b^x\}$ can be determined. We choose to train on the solvable model with $\sigma^z$-type fields 
\begin{align}
H=&\sum \limits_{s} \left(-A_s + e^{- \sum \limits_{i \in s}b^z_i\sigma_i^{z}}\right).
\end{align}
Thus, the network is trained on the inputs $\langle A_s \rangle$, $\langle A_{s'} \rangle$ and $\langle A_s  A_{s'} \rangle$. It is furthermore sufficient to train for a field strength at a specific position (spin), as the measurements for a field at a different position are translationally invariant with respect to the position of the field. In other words, we just choose a single spin to evaluate the expectation values and create the training set by changing the size of the lattice and distribution of the background fields.

\begin{figure}[tb]
\centering\includegraphics[scale=0.72]{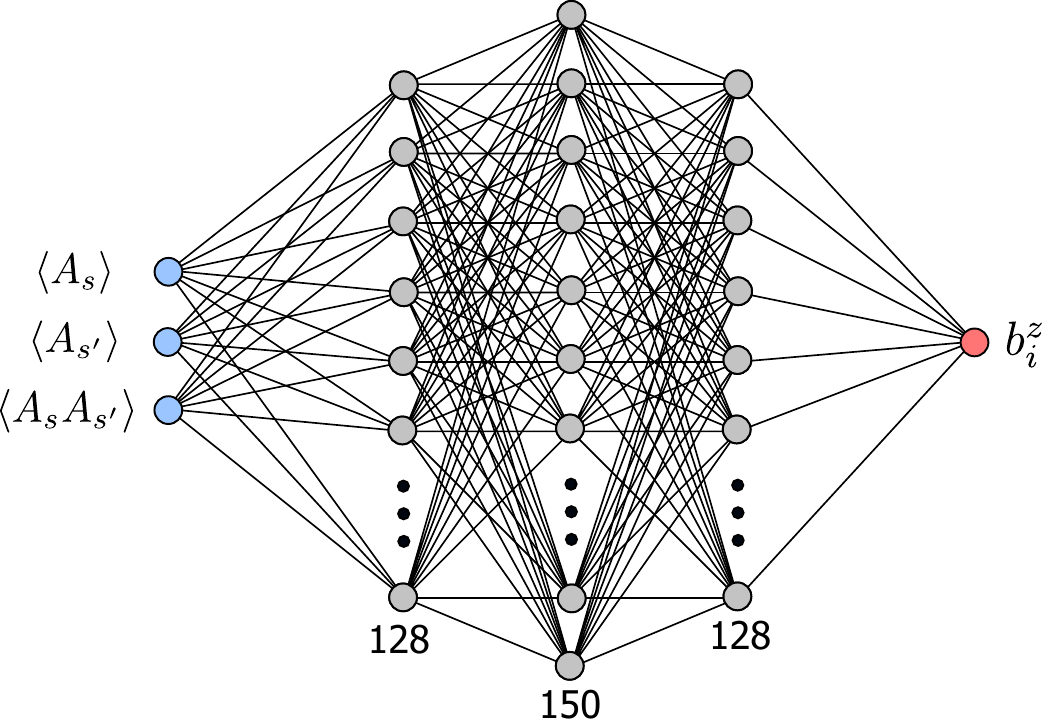}
\caption{The illustrated network consists of 3 input neurons (blue), 3 hidden layers with 128, 150 and 128 neurons and one output neuron (red) to predict the absolute value of the field strength. The same network can be used to estimate the absolute value of $b^ x_i$ with given inputs $\langle B_p \rangle$, $\langle B_{p'} \rangle$ and $\langle B_p B_{p'} \rangle$. }
\label{fig::network}
\end{figure}

We design a neural network with three input neurons, three hidden layers and one output neuron.  The network architecture is depicted in Fig.~\ref{fig::network}. As we estimate a continuous parameter, we calculate the loss function as mean squared distance between the estimated value $b_{\rm output}$ and the correct field strength $b_{\rm label}$
\begin{align}
{C}_{\rm MS}=\frac{||\vec{b}_{{\rm label}}-\vec{b}_{{\rm output}}||^2}{{\rm BS}},
\end{align}
where the vector $\vec{b}_{\rm output/label}$ corresponds to the field vector of the network output and labeled data respectively. The batch size is denoted by ${\rm BS}$.

The network is trained separately for each lattice size $k$, as the stabilizer expectation values vary with lattice size. Training data is generated by calculating the vertex stabilizer expectation values via Monte Carlo sampling for sufficiently distinct field configurations $\{b^z_i\}$, such that the network learns to neglect the contribution in the expectation values from fields we do not aim to determine. More concretely, every field strength is taken from a uniformly random distribution in the interval $[-b_{\rm max},b_{\rm max}]$ and the label is chosen to be the absolute value of the field strength at position $i$.  The maximal value of the field strength $b_{\rm max}=1.7$ is selected such that field configurations are included inducing a probability of a single qubit error after projection into the stabilizer eigenspace larger than the error threshold of standard decoders. We use a traning set of $7450$ examples. The size of the training set is identical for all lattice sizes. We use this restriction in order to be able to analyze scaling behavior (see Fig.~\ref{fig:error+hamil}).

\begin{figure}[tbp]
\centering\includegraphics[scale=1.0]{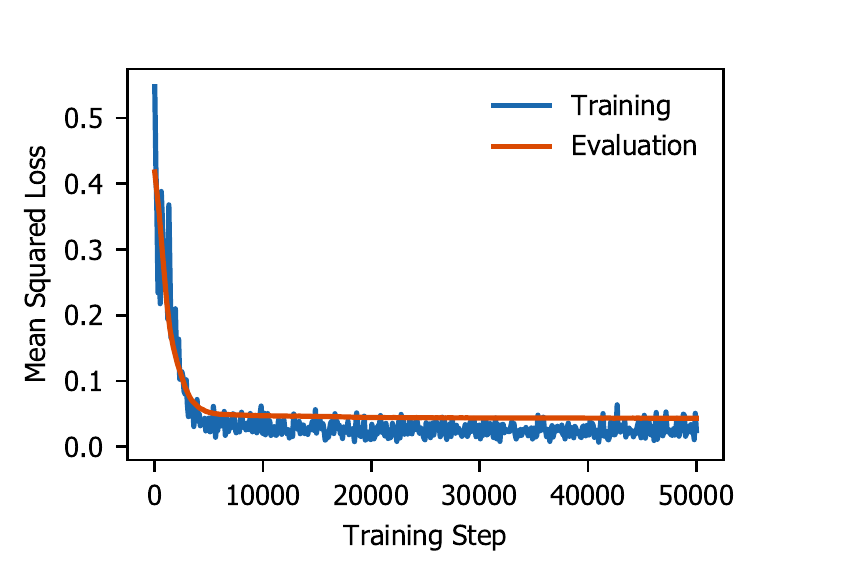}
\caption{Training loss (blue) and evaluation loss (red) calculated as mean squared loss against number of training steps for a lattice of length $k=16$.}
\label{fig::trainevalloss}
\end{figure}

In Fig.~\ref{fig::trainevalloss}, we show that training and evaluation loss (50 evaluation examples) both converge after less than $10^4$ training steps. The training results for the lattice length $k=16$ are plotted. We consider statistical uncertainty in the input expectation values as a main source for the non-zero loss the training converges to, which is similar for all tested lattice sizes. In particular, the neural network has been trained for the lattice sizes $k=3,4,8,12,16,20,24$.

\subsection{Determining the sign}
We use a trained neural network to determine the absolute value of $b^z_i$ and $b^x_i$ by feeding in stabilizer expectation values. These expectation values do not contain sufficient information to determine the sign of $b^z_i$ and $b^x_i$.
In order to find the sign, for each spin we additionally measure the expectation values $\langle \sigma_i^z \rangle$ and $\langle \sigma_i^x \rangle$. We show in the following, that the sign of the field strength can be approximated as the sign of the corresponding expectation value $\langle \sigma_i^z \rangle$ and $\langle \sigma_i^x \rangle$.
In particular, we demonstrate this approximation for a model containing fields in both directions with an additional restriction ensuring solvability. Exact diagonalization and application to the complete Hamiltonian learning process for the analytically non-solvable model discussed throughout this work confirms, that this decision mechnism to determine the sign holds approximately also for the generalized model.
Let us examine the restriction added on the model to keep it solvable. For this purpose, we introduce the sets $I_z$ and $I_x$, which denote the set of spins, for which $b^z_i \neq 0$ and $b^x_i \neq 0$, respectively. We set $I_z \cap I_x = \emptyset$. In different words, only a field in one direction can be present on each spin. The Hamiltonian is then given by
\begin{align}
&H= \sum \limits_{s} (-A_s + e^{- \sum \limits_{i \in s\cap I_z}b^z_i\sigma_i^{z}}) \\ \nonumber
&\ \ \ \ \ \ \ \ +  \sum \limits_{p}(-B_p+e^{- \sum \limits_{i \in p\cap I_x}b^x_i\sigma_i^{x}}).
\end{align}
It can straightforwardly be verified that the un-normalized ground state is of the form
\begin{align}
|\psi\rangle&=e^{\frac{1}{2}\sum \limits_{i \in I_x} b^x_i \sigma_i^x}e^{\frac{1}{2}\sum \limits_{i \in I_z} b^z_i \sigma_i^z}|{\rm GS}\rangle _{\ssm TC}.
\label{eq::psixzalpha}
\end{align}
As we aim to determine a sign, we can ignore overall factors. Hence, we will not discuss normalization here. We calculate the sign of the expectation value $\langle \sigma_i^z \rangle$ measured on the introduced ground state with the restriction $I_z \cap I_x = \emptyset$
\begin{align}
&{\rm sgn} (\langle \sigma_i^z \rangle)= {\rm sgn} ( \langle \Psi|\sigma_i^z| \Psi\rangle )\nonumber \\
&= {\rm sgn}\big(e^{b_i^z}\sum \limits_{g\in G_1 \subset G} e^{\sum \limits_{j, j\neq i} b_j^z \sigma_j^z(g)}  \nonumber \\
&\ \ \ \ -e^{-b_i^z}\sum \limits_{g\in G_2 \subset G} e^{\sum \limits_{j, j\neq i} b_j^z \sigma_j^z(g)} \big) \nonumber \\
&={\rm sgn}(e^{b_i^z}A_1-e^{-b_i^z}A_2).
\label{eq::sigmazexpv}
\end{align}
Any overall factors occurring during the calculation are eliminated by the sign function ${\rm sgn}$. In particular, all dependencies on the parameters $b^x_i$ can be factorized to an overall factor using the condition $I_z \cap I_x = \emptyset$ and therefore vanish.
In addition, we introduced the group $G_1$ as the subgroup $G_1 \subset G$ containing all elements $g$ for which $\sigma_i^z(g)=1$. Analogously, the group $G_2$ is defined such that $\sigma_i^z(g)=-1$ for $g \in G_2$ . The factors $A_1$ and $A_2$ are given by
\begin{align}
A_1= \sum \limits_{g\in G_1 \subset G} e^{\sum \limits_{j, j\neq i} b_j^z \sigma_j^z(g)}, \ \  A_2=\sum \limits_{g\in G_2 \subset G} e^{\sum \limits_{j, j\neq i} b_j^z \sigma_j^z(g)}. \nonumber 
\end{align}
We can relate the sign of the expectation value $\langle \sigma_i^z \rangle$ to the sign of the parameter $b^z_i$ by introducing the approximation $A_1 \approx A_2$\footnote{The two factors are not exactly equal, as the spin $i$ is flipped as part of a vertex and not independently. As a consequence, the value of $\sigma_i^z(g)$ is correlated with the values of the remaining spins in the two adjacent vertices.}. In addition, we make use of the fact that $A_1, A_2 >0$. Combining the two relations, we arrive at
\begin{align}
{\rm sgn} \langle \sigma_i^z \rangle \approx {\rm sgn} (e^{b_i^z}-e^{-b_i^z}), \nonumber
\end{align}
where we dropped the overall factor $A_1$.
Henceforth, the relations
\begin{align}
\begin{cases} b^z_i>0  &\text{if} \ \ \langle \sigma_i^z \rangle>0 \\
b^z_i<0  &\text{if} \ \ \langle \sigma_i^z \rangle<0
\end{cases} \nonumber
\end{align}
hold approximately. Symmetry implies the analogous relations for the parameter $b^x_i$ and the expectation value  $\langle \sigma_i^x \rangle$.

We can summarize the Hamiltonian learning process in two steps. In the first step, the stabilizer expectation values  $\langle A_s \rangle$, $\langle A_{s'} \rangle$, $\langle A_s  A_{s'} \rangle$ and $\langle B_p \rangle$, $\langle B_{p'} \rangle$, $\langle B_p  B_{p'} \rangle$ are measured for each spin and the results fed into the trained neural network, which outputs the absolute values of the field configurations. In the second step, $\langle \sigma_i^z \rangle$ and $\langle \sigma_i^x \rangle$ are measured for each spin and the signs of the parameter determined correspondingly. With the obtained field strengths, the system can be corrected towards the pure toric code. The process can be iterated to achieve optimal results.

\section{Error measures}
\label{app:errormeasures}

\subsection{Probability of a single qubit error}
\label{subsection::errorrate}
On the level of quantum states, we have introduced as an error measure the probability that a single qubit flips its spin or phase after projection into the stabilizer eigenspace. In particular, this probability allows to assess the improvement our introduced method can in principle bring to quantum error correction. More concretely, if the calculated probability is reduced below the decoder-specific threshold, the performance of a standard decoder applied to the system is significantly improved.
Let us make two clarifications to refine the relation to quantum error correction.
The single qubit error threshold is usually defined with respect to the probability of a single qubit error {\it per correction cycle} \cite{FowlerWhiteside2012, BravyiDuclos2012, DennisKitaev2002,FoselTighineanu2018, SwekeKesselring2018}, where a correction cycle corresponds to measurements of stabilizers and applied single-qubit gates. Here, we examine a single cycle and hence, it is sufficient to calculate the probability of a single qubit error. In addition, the examined disorder throughout this work is not of the form of single qubit spin- or phase flip errors. Nevertheless, any disorder is translated to spin- and phase flip errors by projective measurements of the stabilizers. Consequently, we explain in this section how to calculate the single qubit error probability after projecting into the toric eigenspace.

Instead of simulating the projective measurements and ``counting'' the single qubit errors, we calculate the single qubit error probability in a way that does not rely on full quantum simulation and is hence also accessible for large lattice sizes.

We determine the probability of a physical phase-flip (spin-flip), by calculating as first step the probability for an arbitrary vertex (plaquette) to be flipped and deducing the single qubit error probability.
We explain the calculation on the example of vertex flips and single qubit phase-flip errors. It can be repeated analogously for the case of plaquette flips and single qubit bit-flip errors.

\paragraph{Calculating the stabilizer flip rate}
We deduce the vertex stabilizer flip rate using the expectation values $\langle A_s \rangle$, which have already been calculated during the Hamiltonian learning procedure. Every vertex has the eigenvalues $\{\pm 1\}$. Thus, we can make an approximation and re-write a vertex expectation value as
\begin{equation}
\langle A_s \rangle = -p_s+(1-p_s).
\end{equation}
Here, the probability of the vertex $A_s$ to be flipped (projected into an eigenstate with eigenvalue $-1$) is given by $p_s$. We invert the relation to find the probability $p_s$
\begin{equation}
p_s=\frac{1-\langle A_s \rangle}{2}.
\end{equation}
To obtain an average vertex flip rate $p$, we average over all vertices
\begin{equation}
p:=\frac{1}{k^2}\sum \limits_{s} p_s=\frac{1-\frac{1}{k^2}\sum \limits_{s}\langle A_s \rangle}{2}.
\end{equation}

\begin{figure}[tb] 
 \includegraphics[scale=1]{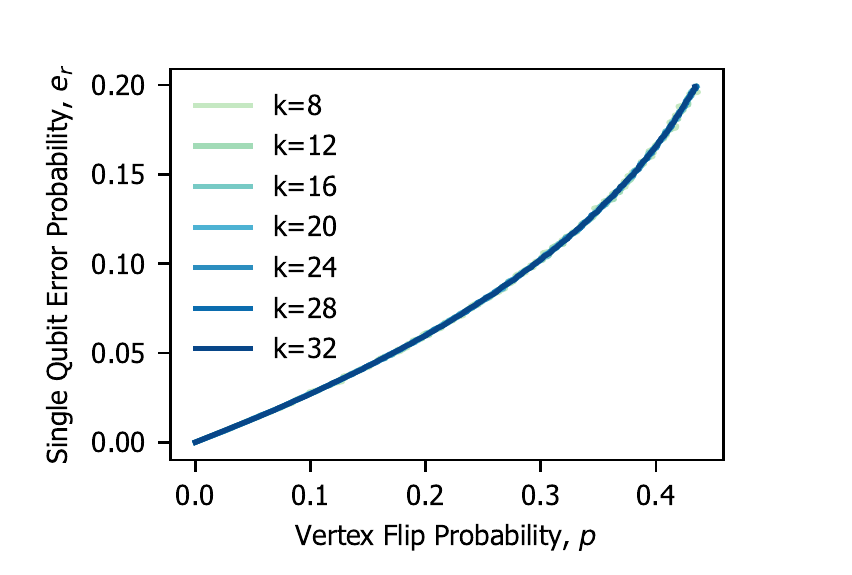}  
\caption{The relation between single qubit error rate $e_r$ on the x-axis and vertex flip rate $p$ on the y-axis obtained by numerical sampling is shown for the lattice sizes $k=8,12,16,20,24,28,32$.}
\label{fig::vertexer_latticesizes}
\end{figure}

  \begin{figure}
  \includegraphics[scale=1]{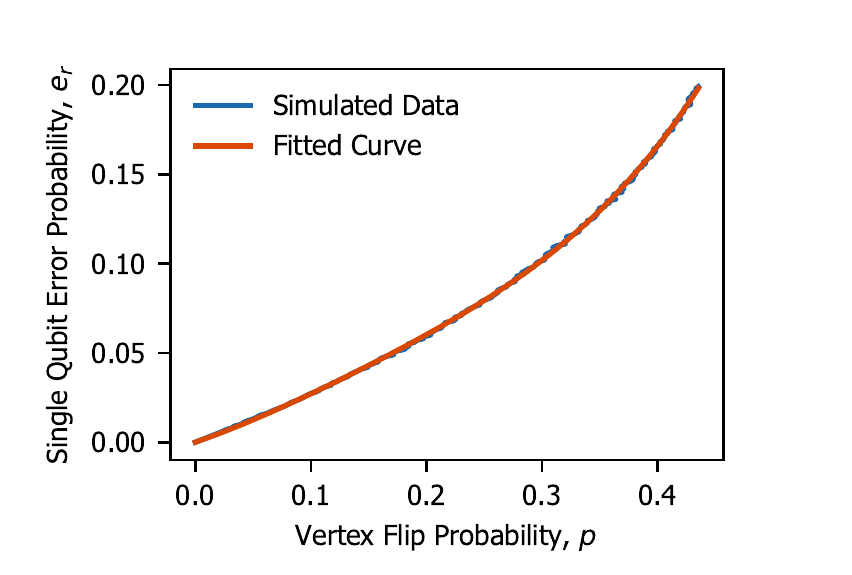}
     \label{fig::vertexer_curvefit} 
  \caption{The numerically obtained relation between single qubit error rate $e_r$ and vertex flip rate $p$ for $k=32$ are plotted together with the fitted curve (see text for details).} 
\label{fig::vertexer}
\end{figure}

\paragraph{Calculating the error rate}
We now understand how to obtain the probability of a vertex flip. We proceed by using the vertex flip probability to deduce the probability of a single qubit phase flip $e_r$ as follows.

We calculate the vertex flip rate $p$ for different probabilities of a single qubit phase flip  by numerically sampling. In particular, we start with a lattice where no phase is flipped and flip each phase with probability $e_r$. The vertex flip probability $p$ is obtained by repeating the procedure, counting the number of flipped vertices and averaging.
We arrive at a function $p(e_r)$.
The inverse function $e_r(p)$ and as a result the probability of a single qubit error can be found by numerical regression.
 
The function $p(e_r)$ is independent of lattice size in the parameter range we are considering here. We verified this by calculating the vertex flip probability for the single qubit phase flip probabilities $e_r\in [0,0.2]$ for the lattice sizes $k=8,12,16,20,24,28,32$, where the total number of spins is equal to $2k^2$ (see Fig.~\ref{fig::vertexer_latticesizes}).

We can conclude that it is sufficient to find a numerical fit for the function of one lattice size.
To minimize uncertainties caused by the statistical error of the simulation, we pick the largest simulated lattice size, $k=32$, to fit the curve.
We use an ansatz of a polynomial of degree $4$ and minimize the mean squared distance.
The curve is fitted up to a mean squared loss of $l_{\rm ms}\approx 10^{-7}$ by the function
\begin{equation}
e_r(p)= 0.2187p+  0.72419p^2 -2.5398p^3+  4.90118p^4.
\end{equation}
Both numerically simulated data and the fit are shown in Fig.~\ref{fig::vertexer}. 
The obtained function returns the probability of a single qubit phase flip given the vertex stabilizer expectation values and similarly applies to single qubit bit-flip errors given plaquette expectation values.
Full quantum simulation for $k=3$ confirms these results.

\subsection{Measure for Hamiltonian error}
\label{subsection::Hamilerr}
An error measure on the level of Hamiltonian learning is given by the reconstruction error $\Delta_H$ \cite{BaireyArad2019}. In this section, we detail the calculation of this Hamiltonian error for the disordered toric code model.
As a reminder, $\Delta_H$ is given by (\ref{eq::delta2})
\begin{align}
\label{eq::delta2_app}
\Delta_H=||\hat{c}_{\rm true}-\hat{c}_{\rm recovered}||_2,
\end{align}
where the vector $\hat{c}$ corresponds to the normalized coefficient vector $\hat{c}=\frac{\vec{c}}{||c||}$. 
The coefficients $c$ are obtained by expanding the Hamiltonian in a suitable basis $S$
\begin{align}
H=\sum \limits_{m} c_m S_m,
\end{align}
where all parameters we estimate are contained in the coefficients, not in the basis.
Let us expand the disordered toric Hamiltonian in an operator basis
\begin{align}
\label{eq::Hamexpansion}
&H=\sum \limits_{s} (-A_s + e^{-\sum \limits_{i \in s}b^z_i\sigma_i^{z}})+  \sum \limits_{p}(-B_p+e^{-\sum \limits_{i \in p} b^x_i\sigma_i^{x}}) \nonumber \\
&=\mathlarger\sum \limits_{s}\big( -\mathop{\prod_{i \in s}}\sigma_i^x+\mathop{\prod_{i \in s}}\cosh(b^z_i)){1} \nonumber \\ 
&-\sum \limits_{i \in s} \sinh(b^z_i)\prod \limits_{j \in s, j \neq i} \cosh(b^z_j) \sigma_i^{z} \nonumber \\
&+\mathop{\sum_{i < j \in s}} \sinh(b^z_i)\sinh(b^z_j) \mathop{\prod_{l<m\in s}} \limits_{l\neq i \neq m \neq j} \cosh(b^z_l) \cosh(b_m) \sigma_i^{z}\sigma_j^{z} \nonumber \\
&-\mathop{\sum_{j<k<l \in s}} \limits_{i \neq j \neq k \neq l \in s}\cosh(b^z_i)\sinh(b^z_j)\sinh(b^z_k) \sinh(b^z_l)\sigma_j^{z}\sigma_k^{z}\sigma_l^{z} \nonumber 
\\ 
&+\prod \limits_{i \in s} \sinh(b^z_i)\sigma_i^z\big)  \nonumber\\
&+(s\leftrightarrow p, x \leftrightarrow z, b^z_i \leftrightarrow b^x_i).
\end{align}
Each term in the expansion of the Hamiltonian is a product of the form $c_i S_i$, with $c_i$ being a coefficient and $S_i$ an operator. The dependence on the parameters $b^z_i$ and $b^x_i$ is fully contained in the coefficients $c_i$, the operator basis is {\it independent} of the parameters. More specifically, the operator basis resulting from the Hamiltonian expansion is a combination of the unit matrix and products of Pauli matrices. We illustrate the separation into coefficients and basis by re-writing all occurring operators as $S_i$, an (arbitrarily) numbered basis element. The expansion of the Hamiltonian can be divided in two parts: a sum over all vertices $s$ and a sum over all plaquettes $p$. For a simpler basis numbering, all basis elements (operators) occuring in the sum over vertices are denoted by $S_{s,i}$ and the operators in the sum over plaquettes are denoted by $S_{p,i}$. Inserting the basis elements yields

\begin{align}
&H=
\mathlarger\sum \limits_{s}\big( -1\cdot S_{s,0}+\mathop{\prod_{i \in s}}\cosh(b^z_i))S_{s,1} \nonumber \\
&-\sum \limits_{i \in s} \sinh(b^z_i)\prod \limits_{j \in s, j \neq i} \cosh(b^z_j)S_{s,1+i}  \nonumber \\
&+\mathop{\sum_{i < j \in s}} \sinh(b^z_i)\sinh(b^z_j) \mathop{\prod_{l<m\in s}} \limits_{l\neq i \neq m \neq j} \cosh(b^z_l) \cosh(b^z_m) S_{s,5+4i+j} \nonumber \\
&-\mathop{\sum_{j<k<l \in s}} \limits_{i \neq j \neq k \neq l \in s}\cosh(b^z_i)\sinh(b^z_j)\sinh(b^z_k) \sinh(b^z_l)S_{s,11+i} \nonumber \\
&+\prod \limits_{i \in s} \sinh(b^z_i)S_{s,15+i}\big)  \nonumber\\
&+(s\leftrightarrow p, x \leftrightarrow z, b^z_i \leftrightarrow b^x_i) \\
&=\mathlarger\sum \limits_{s,i}\ c_{s,i}S_{s,i}+\mathlarger\sum \limits_{p,i} c_{p,i}S_{p,i}=\sum \limits_{m} c_m S_m.
\end{align}
The field parameters $b^z_i$ and $b^x_i$ only occur in the coefficients $c_m$. 
The operator basis $\{S_m\}$ is the joined set of the basis elements contained in $\{S_{s,i}\}$ and in $\{S_{p,i}\}$. Similarly, the coefficients $\{c_m\}$ are given by $\{c_m\}$=$\{c_{s,i}\} \cup \{c_{p,i}\}$ and can be read off easily.
Having hereby deduced the coefficients from the field parameters $\{b^z_i\}$ and $\{b^x_i\}$, we can calculate the Hamiltonian error $\Delta_H$ using Eq. (\ref{eq::delta2_app}).

\end{appendix}

\bibliographystyle{unsrt}
\bibliography{toric_new.bib}

\end{document}